\documentclass{ws-procs9x6}
\usepackage{graphicx}
\begin{document}
\newcommand{\Eq}[1]{Eq.~(\ref{#1})}
\newcommand{\Fig}[1]{Fig.~\ref{#1}}

\title{The Thermal Glueball at Finite Temperature in SU(3) Anisotropic
Lattice  QCD\footnote{\uppercase{T}he numerical  calculation  has been
done     on      \uppercase{NEC     SX-5}     at     \uppercase{O}saka
\uppercase{U}niversity.}}
\author{Noriyoshi ISHII}
\address{Radiation Lab.,
RIKEN(The Institute of Physical and Chemical Research)\\
2-1 Hirosawa, Wako-shi, Saitama 351-0198, JAPAN\\
E-mail: ishii@rarfaxp.riken.go.jp
}
\author{Hideo  SUGANUMA\footnote{supported  by  \uppercase{G}rant  for
\uppercase{S}cientific      \uppercase{R}esearch      (\uppercase{N}o.
12640274) from    \uppercase{M}inistry    of    \uppercase{E}ducation,
\uppercase{C}ulture,  \uppercase{S}cience  and \uppercase{T}echnology,
\uppercase{J}apan.}}
\address{Faculty of Science, Tokyo Institute of Technology,\\
2-12-1 Ohkayama, Meguro, Tokyo 152-8552, JAPAN
%E-mail: suganuma@th.phys.titech.ac.jp
}
\author{Hideo    MATSUFURU\footnote{supported   by   \uppercase{J}apan
\uppercase{S}ociety      for     the      \uppercase{P}romotion     of
\uppercase{S}cience for \uppercase{Y}oung \uppercase{S}cientists}}
\address{Yukawa Institute for Theoretical Physics, Kyoto University,\\
Kitashirakawa-Oiwake, Sakyo, Kyoto 606-8502, JAPAN
%E-mail: matufuru@rcnp.osaka-u.ac.jp
}


\maketitle

\abstracts{ The  thermal glueball is studied at  finite temperature by
using SU(3) anisotropic  lattice QCD with $\beta_{\rm{lat}}=6.25$, the
renormalized anisotropy  $\xi\equiv a_s/a_t =  4$ over the  lattice of
the size  $20^3\times N_t$ with  various $N_t$ at the  quenched level.
While  the  narrow  peak  ansatz  leads to  the  significant  polemass
reduction  of  about 300  MeV  near  the  critical temperature  $T_c$,
Breit-Wigner  ansatz   which  can  take  into   account  the  possible
appearance of the thermal width leads to the significant thermal width
broadening of about 300 MeV with a modest reduction of the peak center
of about 100 MeV.}

At finite  temperature/density, the QCD vacuum  changes its structure.
This is the case even in  the confinement phase, where one expects the
reduction in  the string  tension and the  partial restoration  of the
spontaneous chiral symmetry  breaking.  Since hadrons are relativistic
bound states of quarks and gluons,  it would be natural to expect that
the  changes  in  the QCD  vacuum  leads  to  the changes  in  various
properties of hadrons.
%% Hadrons are relativistic bound states  of quarks and gluons. Hence, at
%% finite temperature/density  even in the confinement  phase, we expect
%% that they  change their properties  reflecting the changes of  the QCD
%% vacuum such  as the  reduction of the  string tension and  the partial
%% chiral restoration.
In fact,  near the critical  temperature $T_c$, a number  of effective
models predict  the (pole)mass reductions  of various hadrons  of more
than a few hundred MeV \cite{hatsuda1,hashimoto,hatsuda2,ichie}, which
would  serve  as important  precritical  phenomena  of  the QCD  phase
transition.    Recently,   motivated   by  these   studies,   quenched
anisotropic lattice  QCD \cite{klassen} has  been used to  measure the
polemass     of    various     hadrons    at     finite    temperature
\cite{taro,umeda,ishii1},  reporting the  profound  results that,  for
both the  light and heavy $q\bar{q}$-mesons, no  significant change is
observed  below  $T_c$,  while,  for  the  glueball,  the  significant
polemass reduction of about 300 MeV is observed near $T_c$.
In  all of  these  studies, the  narrowness  of the  peak is  assumed.
However, since at  $T\neq 0$, each bound state  peak acquires a finite
thermal  width  through  the  interaction  with the  heatbath,  it  is
desirable to  respect the  existence of the  thermal width.   Here, we
report an advanced analysis of  the thermal $0^{++}$ glueball based on
SU(3) quenched anisotropic lattice  QCD taking into account the effect
of the thermal width.

Generally, to extract physical observables  such as mass and width, we
have  to  resort  to  the  spectral representation  of  the  two-point
correlator  $G(t)  \equiv  Z(\beta)^{-1} \mbox{tr}\left\{e^{-\beta  H}
\phi(t) \phi(0)\right\} $ as
\begin{equation}
  G(t) =
  \int_{-\infty}^{\infty} {d\omega\over 2\pi}
  \frac{\rho(\omega)}{2\sinh(\beta\omega/2)}
  \cosh\left(\omega(\beta/2 - t)\right),
\label{spec}
\end{equation}
where $H$  is the QCD Hamiltonian,  $Z\equiv \mbox{tr}(e^{-\beta H})$,
$\phi(t) \equiv  e^{tH}\phi(0)e^{-tH}$ is the  zero-momentum projected
glueball  operator  in  the  imaginary-time  Heisenberg  picture,  and
$\rho(\omega)$  is  the  spectral  function. Appropriate  smearing  on
$\phi(t)$ is understood to maximize the overlap to the glueball state.
To  extract the physical  observables, we  parameterize $\rho(\omega)$
and  use  \Eq{spec}  to  fit  $G(t)$ generated  by  lattice  QCD.   In
Refs.~[6,7,8],  the   narrow-peak  ansatz  has   been  adopted,  where
$\rho(\omega)$  is  parameterized  as  $\rho(\omega)  =  2\pi  A\left(
\delta(\omega - m) - \delta(\omega + m) \right) + \cdots$ with the two
fit  parameters $A$  and  $m$  corresponding to  the  overlap and  the
polemass, respectively.

To respect  the thermal  width, we recall  that $\rho(\omega)$  is the
imaginary  part of  the retarded  Green function  $G_R(\omega)$, i.e.,
$\rho(\omega)  =  -  2\mbox{Im}\left(G_R(\omega)\right)$.   At  $T=0$,
bound  state   poles  of  $G_R(\omega)$   are  located  on   the  real
$\omega$-axis. With  increasing $T$, they  begin to move off  the real
axis  into the  complex $\omega$-plane.   Thus the  influence  of each
complex   pole  in   $\rho(\omega)$   amounts  to   a  Lorentzian   as
$\rho(\omega)  =  2\pi A\left(  \delta_{\Gamma}(\omega  - \omega_0)  -
\delta_{\Gamma}(\omega  +  \omega_0)  \right)  + \cdots$,  where  $A$,
$\omega_0$ and  $\Gamma$ are used  as fit parameters  corresponding to
the  overlap,   the  center  and  the   thermal  width,  respectively.
$\delta_{\epsilon}(x) \equiv  \frac1{\pi} \mbox{Im} \left(  \frac1{x -
i\epsilon}\right)$ denotes a smeared  delta function.  We refer to the
corresponding fit function as the Breit-Wigner type.

\begin{figure}[ht]
\begin{center}
\includegraphics[angle=-90,scale=0.16]{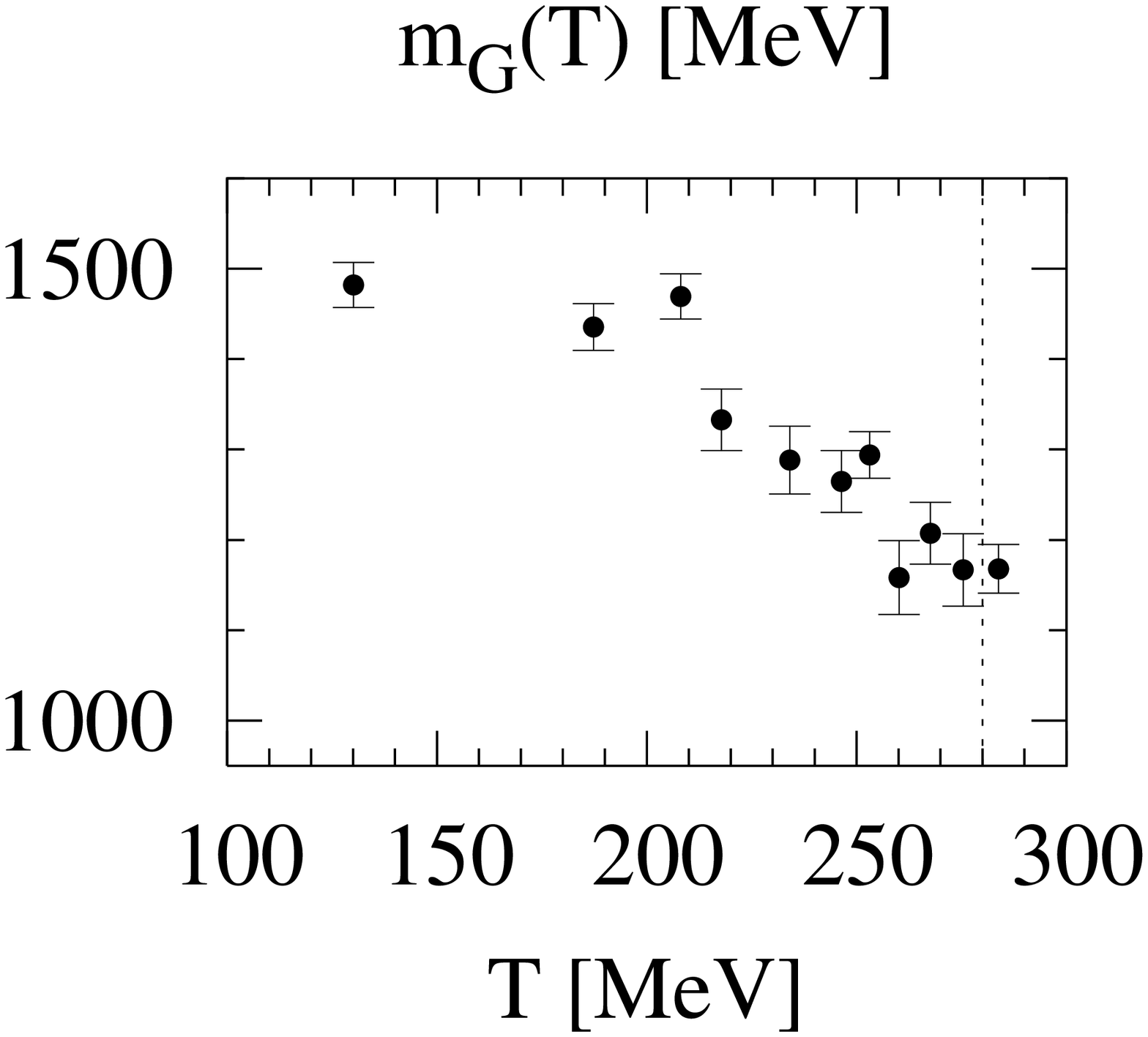}
\hspace{-3.5em}
\includegraphics[angle=-90,scale=0.16]{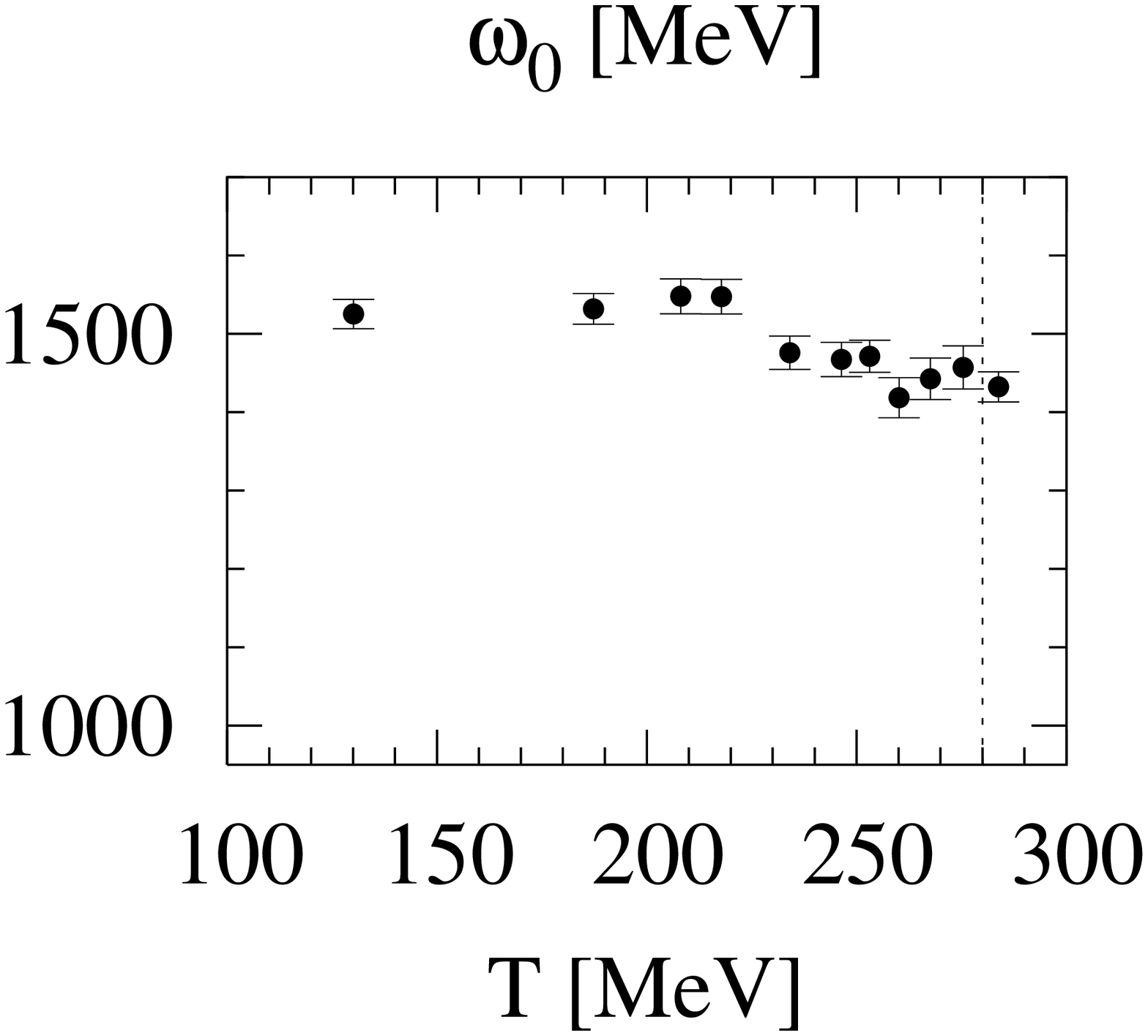}
\hspace{-3.5em}
\includegraphics[angle=-90,scale=0.16]{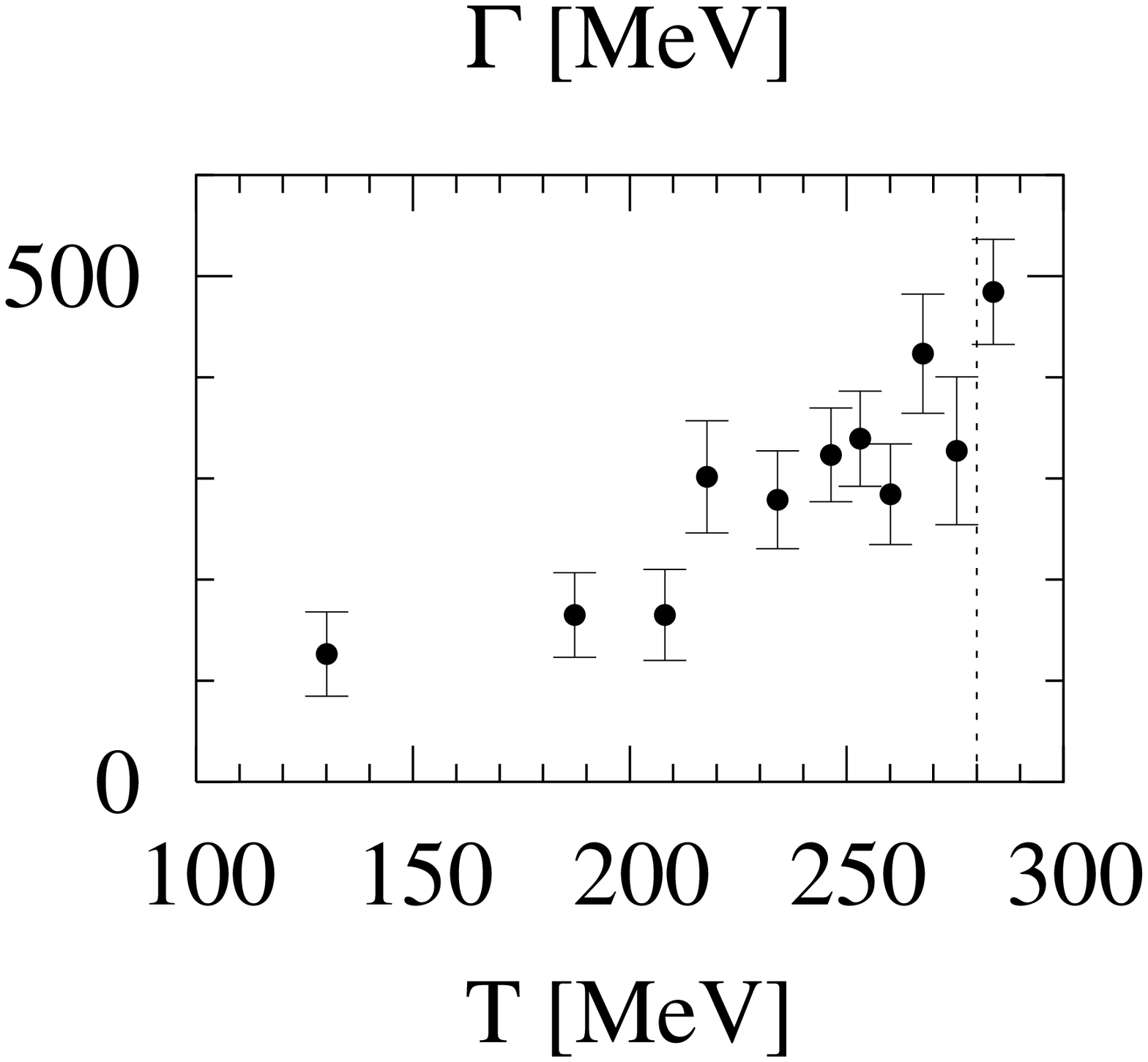}
\end{center}
\caption{The polemass $m_{\rm G}(T)$  from the narrow-peak ansatz, the
center  $\omega_0(T)$  and  the  thermal width  $\Gamma(T)$  from  the
Breit-Wigner ansatz.  The vertical  dotted lines indicate the critical
temperature   $T_c\simeq   280$   MeV.   Appropriate   smearings   are
understood.}
\label{figure}
\end{figure}

The numerical results  are shown in \Fig{figure}. We  use 5000 to 9900
gauge configurations generated by  the Wilson action with $\beta=6.25$
and the renormalized anisotropy  $\xi\equiv a_s/a_t = 4$.
Whereas  the  narrow-peak ansatz  indicates  the significant  polemass
reduction of  about $300$ MeV,  the Breit-Wigner ansatz  indicates the
significant thermal  width broadening  of more than  $300$ MeV  with a
modest reduction in  the peak center.  These two  analysises thus lead
to the  two different  physical implications.  Note  that, due  to the
biased factor  ``$\sinh(\beta\omega/2)$'' in \Eq{spec}  which enhances
the  smaller  $\omega$  region  of $\rho(\omega)$,  thermal  width  is
effectively seen  as the reduced  polemass in the  narrow-peak ansatz.
Hence,  in  the case  of  the glueball,  the  thermal  effect is  most
probably the thermal width broadening.

To  summarize,   we  have  studied  the  thermal   glueball  by  using
anisotropic  lattice QCD  at  quenched level.   While the  narrow-peak
ansatz has lead to the  significant polemass reduction near $T_c$, the
Breit-Wigner  analysis  has  lead  to the  significant  thermal  width
broadening  with a  modest reduction  in the  peak center.   Since the
thermal width  can be effectively  seen as the polemass  reduction, we
have  concluded  that the  thermal  effect  on  the glueball  is  most
probably the thermal width broadening.

\end{document}